\documentclass[aps,prl,floatfix,twocolumn,showpacs,nofitinbib,hyperref=pdftex,citeautoscript]{revtex4-1}
\usepackage{epsfig}
\usepackage{amsmath}
\usepackage{amssymb}
\usepackage{amsfonts}
\usepackage{ dsfont }
\usepackage{ comment,color }
\usepackage{lipsum}
\usepackage{hyperref}

\hypersetup{colorlinks=true,linkcolor=blue,anchorcolor=blue,citecolor=blue,filecolor=blue,urlcolor=blue,bookmarksnumbered=true,pdfview=FitB}

\newcommand{\bk}{{{\bf{k}}}}

\newcommand{\beqa}{\begin{eqnarray}}
\newcommand{\eeqa}{\end{eqnarray}}

\newcommand{\ua}{\uparrow}
\newcommand{\da}{\downarrow}

\begin{document}

\hsize\textwidth\columnwidth\hsize\csname@twocolumnfalse\endcsname

\title{Andreev or Majorana, Cooper finds out} 
\author{Constantin Schrade and Liang Fu}
\affiliation{Department of Physics, Massachusetts Institute of Technology, 77 Massachusetts Ave., Cambridge, MA 02139}
\date{\today}

\vskip1.5truecm
\begin{abstract}
We study a Cooper pair transistor realized by a mesoscopic superconductor island that couples to a pair of $s$-wave superconducting leads. For a trivial island, the critical supercurrent between the leads exhibits a well-known $2e$-periodicity in the island-gate charge. Here, we show that for an island with spatially separated zero-energy Majorana or Andreev bound states the periodicity of the \textit{magnitude} of the critical supercurrent transitions to $1e$ in the island-gate charge. Moreover, for Andreev bound states the current-phase relation displays a \textit{sign reversal} when the parity of the charge ground state of the island changes between even and odd. Notably, for Majorana bound states the same sign reversal does not occur. Our results highlight the relevance of measuring the \textit{full current-phase relation} of a Cooper pair transistor for clarifying the nature of zero-energy bound states in candidate systems for topological superconductors and provide an initial step towards integrating Majorana qubits in superconducting circuits.  
\end{abstract}

% 74.50.+r: Tunneling phenomena; Josephson effects
% 85.25.Cp Josephson devices
% 71.10.Pm: Fermions in reduced dimensions

\maketitle

Topological superconductors (TSCs) hosting spatially separated Majorana bound states (MBSs) form a key component of robust quantum computing architectures  \cite{bib:Alicea2012,bib:Beenakker2013,bib:Lutchyn2017,bib:Aguado2017,bib:Bravyi2010,bib:Vijay2015,bib:Vijay2016,bib:Landau2016,bib:Plugge2016,bib:Hoffman2016,bib:Vijay2016_2,bib:Plugge2017,bib:Karzig2016,bib:Schrade2018_2,bib:Schrade2018_3}. Proposed realizations of TSCs comprise superconductor (SC) - semiconductor nanowires under strong magnetic fields \cite{bib:Lutchyn2010,bib:Oreg2010,bib:Mourik2012,bib:Das2012,bib:Rokhinson2012,bib:Deng2013}, magnetic atom chains on a SC \cite{bib:Klinovaja2013,bib:Braunecker2013,bib:Pientka2013,bib:Nadj-Perge2013,bib:Nadj-Perge2014,bib:Ruby2015,bib:Pawlak2016} as well as vortex cores in SC-topological insulator devices \cite{bib:Fu2008,bib:Xu2015,bib:Sun2016}. Most notably, in these candidate platforms, the emergence of a zero-bias conductance peak has been perceived as a first step towards verifying the existence of MBSs \cite{bib:Law2009,bib:Flensberg2010,bib:Sau2010}. However, zero-bias conductance peak measurements have difficulties in differentiating a zero-energy MBS from zero-energy Andreev bound states (ABSs) that can also appear in the systems listed above \cite{bib:Lee2012,bib:Kells2012,bib:Prada2012,bib:Avila2018}. Hence, we need a more refined diagnostic tool to discriminate between MBSs and ABSs.

\begin{figure}[!t] \centering
\includegraphics[width=1\linewidth] {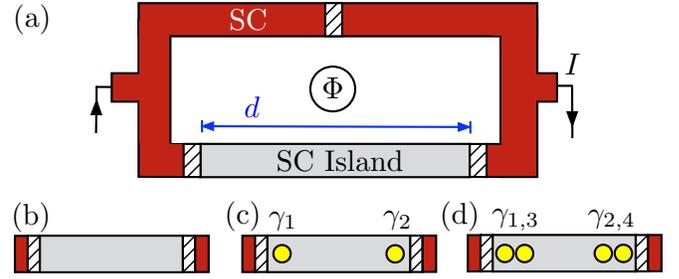}
\caption{(Color online)
(a) A CPT realized by mesoscopic SC island (gray) of length $d$ that weakly couples to a pair of $s$-wave SC leads (red). 
The two SC leads also couple directly to each other forming a SQUID loop with magnetic flux $\Phi$. 
(b) Trivial CPT. (c) MST with two spatially separated MBSs $\gamma_{1},\gamma_{2}$ (yellow). (d) AST with two spatially separated ABSs that are decomposable into four MBSs $\gamma_{1},\gamma_{2},\gamma_{3},\gamma_{4}$ (yellow).}\label{fig:1}
\end{figure}

With the goal of creating such a more refined diagnostic tool, we  revisit a well-established superconducting circuit element: The Cooper pair transistor (CPT); realizable by a mesoscopic SC island coupled to $s$-wave SC leads, see Fig.~\ref{fig:1}(a). In the trivial regime, shown in Fig.~\ref{fig:1}(b), the properties of CPTs have been experimentally studied for many years \cite{bib:Fulton1989}. Most crucially, the supercurrent across a CPT exhibits, in the absence of quasiparticle poisoning \cite{bib:Aumentado2004,bib:Lutchyn2007,bib:Shaw2008,bib:Court2008}, a characteristic $2e$-periodicity in the island-gate charge \cite{bib:Joyez1994,bib:Yamamoto2006,bib:Ferguson2006,bib:Corlevi2006,bib:Woerkom2015,bib:Veen2018}. 

In this work, we generalize the concept of the CPT in two ways: First, to the ``Majorana superconducting transistor" (MST) where the SC island hosts two spatially separated MBSs, see Fig.~\ref{fig:1}(c). Second, to the ``Andreev superconducting transistor" (AST) where the SC island hosts two spatially separated ABSs that are decomposable into two MBSs each, see Fig.~\ref{fig:1}(d). We find that for both devices, the magnitude of the critical supercurrent exhibits a characteristic $1e$-periodicity in the island-gate charge provided that the MBSs or the ABSs reside close to zero energy. While this signature discriminates the trivial, unpoisoned CPT from both the MST and the AST, the MST and the AST are also distinguishable among themselves:  For the AST, the current-phase relation reverses its sign when the parity of the island-ground state charge changes between even and odd. In contrast, the MST shows no such sign reversal. 

Our findings highlight the significance of measuring the \textit{full current-phase relation} in the MST or AST for understanding the nature of zero energy bound states in TSC candidate systems. In addition, the concept of the MST provides a first step towards integrating a Majorana qubit in an all-superconducting circuit \cite{bib:Schrade2018_2,bib:Schrade2018_3}. Such an integration is attractive as it promises improved protection from quasiparticle poisoning due to the finite SC gap in the circuit and may, therefore, constitute a viable platform for Majorana-based quantum computing.

{\it Setup.}
We study a CPT realized by a mesoscopic SC island that connects to the ground via a capacitor and weakly couples to a pair of $s$-wave SC leads, see Fig.~\ref{fig:1}(a). Within this setup, the two SC leads also couple directly to each other and, in this way, form a superconducting quantum interference device (SQUID). We label the SC leads by $m=\text{L,R}$ and we describe them in terms of a BCS (Bardeen-Cooper-Schrieffer) Hamiltonian,
\begin{equation}
\label{Eq1}
H_{\text{SC}}=\sum_{m=\text{L,R}}\sum_{\bk} \Psi_{m,\bk}^\dagger \left(
\xi_{\bk}\eta_{z}+\Delta_{m}\eta_{x}e^{i\varphi_{m}\eta_{z}}
\right)\Psi_{m,\bk}.
\end{equation}
Here, $\Psi_{m,\bk}=(c_{m,\bk\ua},c^{\dag}_{m,-\bk\da})^{T}$ is the Nambu spinor in the $m$-SC lead acting on the Pauli matrices $\eta_{x,y,z}$ with $c_{m,\bk s}$ being the electron annihilation operator for momentum $\bk$ and spin $s=\ua,\da$. The single-particle dispersion is $\xi_{\bk}$. Moreover, the SC gaps and phases are $\Delta_{m}$, $\varphi_{m}$. For simplicity, we will assume that both SC leads have identical SC gaps, $\Delta\equiv\Delta_{\text{L}}=\Delta_{\text{R}}$. 

Since the SC island is of mesoscopic size, it acquires a substantial charging energy that suppresses extrinsic quasiparticle poissoning, 
\begin{equation}
U_{C}(n) = (ne-Q)^{2}/2C.
\end{equation}
Here, $n$ counts the number of electron charges $e$ on the SC island and $Q$ is a gate charge which we can tune  continuously via a gate voltage $V_{g}$ across a capacitor with capacitance $C$. We distinguish two regimes: 

(1) In the \textit{MST regime}, which is depicted in Fig.~\ref{fig:1}(b), the SC island hosts two MBSs $\gamma_{1}, \gamma_{2}$ at opposite terminal points. If the length $d$ of the SC island is comparable to the MBS localization length, the two MBSs $\gamma_{1}, \gamma_{2}$ acquire a finite energy splitting $\varepsilon_{12}$. We model this energy splitting by the Hamiltonian, 
\begin{equation}
H_{\text{MBS}}=i\varepsilon_{12}\gamma_{1}\gamma_{2}.
\end{equation}
By tuning the gate charge $Q/e$ so that the SC island hosts $n_{0}$ electron charges in its ground state, the total fermion parity of the SC island obeys \cite{bib:Fu2010,bib:Xu2010}, 
\begin{equation}
i\gamma_{1}\gamma_{2}=(-1)^{n_{0}}.
\end{equation}
For well-separated MBSs, $\varepsilon_{12}=0$, this constraint reduces the two-fold degenerate ground state 
at zero charging energy to a non-degenerate ground state. 

(2) In the \textit{AST regime}, which is shown in Fig.~\ref{fig:1}(c), the SC island hosts two ABSs at opposite terminal points whose field operators are decomposable into four MBSs $\gamma_{1},\gamma_{2},\gamma_{3},\gamma_{4}$ in total. 
We model the coupling between the four MBSs by the Hamiltonian, 
\begin{equation}
H_{\text{ABS}}=i\sum_{i<j}\varepsilon_{ij}\gamma_{i}\gamma_{j}. 
\end{equation}
Here, $\varepsilon_{ij}$ are coupling constants. Similar to the topological regime, we tune the gate charge $Q/e$ so that the SC island hosts $n_0$ electron charges in its ground state. 
The total fermion parity then satisfies, 
\begin{equation}
\gamma_{1}\gamma_{2}\gamma_{3}\gamma_{4}=(-1)^{n_{0}}.
\label{Constraint}
\end{equation}
For zero-energy ABSs, $\varepsilon_{ij}=0$, this constraint reduces the four-fold degeneracy of the ground state at zero charging energy to a two-fold degeneracy. 

Next, we couple the SC leads to the MBSs on the SC island. The tunnelling Hamiltonian is \cite{bib:Zazunov2012,bib:Zazunov2017,bib:Zazunov2018,bib:Schrade2018_1},
\begin{align}
\label{Eq4}
H_{T}
&=
\sum_{m,i}
\sum_{\bk,s}
\lambda^{s}_{m i} \
c^{\dag}_{m,\bk s}
\gamma_{i}
e^{-i\phi/2}
+
\text{H.c.}
\end{align}
Here, the point-like and complex tunnelling amplitudes $\lambda^{s}_{m i}$ couple the lead fermions in the $m$-SC to the MBSs $\gamma_{i}$ where $i=1,2$ for the MST and $i=1,...,4$ for the AST. Notably, the lead fermions couple to all MBSs due to the finite length $d$ of the SC island \cite{bib:Schuray2018,bib:Prada2018}. In addition, the operators $e^{\pm i\phi/2}$ increase/decrease the total charge of the SC island by one unit, $[n,e^{\pm i\phi/2}]=\pm e^{\pm i\phi/2}$ while the MBS operators $\gamma_{i}$ flip the number parity. Lastly, couplings to above-gap quasiparticles in the SC island are negligible assuming a sufficiently large SC gap in the SC island. Such a large SC gap with no subgap conductance is conceivable in semiconductor nanowires or two-dimensional electron gases proximitized by an Al/NbTi/NbTiN multilayer SC \cite{bib:Drachmann2017}.

In a final step, we connect the SC leads directly to each other via a conventional Josephson junction. We describe the latter by the tunneling Hamiltonian, 
\begin{align}
\label{Eq5}
H_{T,\text{ref}}
&=
\sum_{\bk,s}
\lambda_{\text{ref}} e^{i\pi\Phi/\Phi_{0}}
c^{\dag}_{\text{L},\bk s}
c_{\text{R},\bk s}
+
\text{H.c.}
\end{align}
Here, for simplicity, the point-like tunnelling amplitude $\lambda_{\text{ref}}$ is taken to be real and spin-independent. 
Moreover, $\Phi$ denotes a flux piercing through the SQUID-loop and $\Phi_{0}=e/2h$ is the flux quantum. We have made the inessential assumption that the tunneling is spin-conserving.

In summary, the full Hamiltonian for the MST is given by $H=H_{\text{SC}}+U_{C}(n)+H_{\text{MBS}}+H_{T}+H_{T,\text{ref}}$ and for the AST by $H'=H_{\text{SC}}+U_{C}(n)+H_{\text{ABS}}+H_{T}+H_{T,\text{ref}}$.

{\it Supercurrent in the MST regime.}
We are now in the position to compute the supercurrent due to Cooper pair tunnelling between the SC leads mediated by the MBSs $\gamma_{1}$, $\gamma_{2}$ on the MST. We will focus on nearly-zero-energy MBSs, $\varepsilon_{12}\ll\Delta,U$ with $U\equiv e^{2}/2C$, which is the only relevant case for qubit applications \cite{bib:Bravyi2010,bib:Vijay2015,bib:Vijay2016,bib:Landau2016,bib:Plugge2016,bib:Hoffman2016,bib:Vijay2016_2,bib:Plugge2017,bib:Karzig2016,bib:Schrade2018_2,bib:Schrade2018_3}. 

As a starting point, we note that to second order in the tunnelling amplitudes $\lambda^{s}_{m i}$, the energy gap $\Delta$ of the SC leads suppresses single electron transfer across the SC island. Similarly, the charging energy $U$ of the SC island suppresses Cooper pair transfer between each individual SC lead and the SC island. Consequently, no Josephson coupling between the SC leads arises from second-order processes in the tunnelling amplitudes $\lambda^{s}_{m i}$.

In a next step, we examine fourth-order sequences in the tunnelling amplitudes $\lambda^{s}_{m i}$. Here, we find that the only sequences which generate a finite Josephson coupling involve a Cooper pair moving between  the SC leads by tunnelling \textit{both in and out} of the SC island via the spatially separated MBSs $\gamma_{1}$ and $\gamma_{2}$. Crucially, these sequences are $\propto (\gamma_{1} \gamma_{2})^{2}=1$ and, therefore, independent of the total fermion parity of the SC island. 

We calculate the amplitudes of the relevant sequences in the weak coupling limit, $\pi\nu_{m}|\lambda^{s}_{m i}|^{2},\pi\nu_{m}\lambda_{\text{ref}}^{2}\ll\Delta, U$ with $\nu_{m}$ the normal-state density of states per spin of the $m$-SC at the Fermi energy. We summarize our results by an effective Hamiltonian,
\begin{equation}
H_{\text{eff}}=
-
J_{\text{ref}}\cos\varphi_{\text{ref}}
-
J
\cos\varphi\label{Heff1},
\end{equation}
where we disregard all contributions that are independent of the SC phases and do not add to the supercurrent. 

The first term in the effective Hamiltonian describes the Josephson junction that directly couples the two SC leads.
Here, $J_{\text{ref}}\sim \lambda_{\text{ref}}^{2}/\Delta$ is the corresponding Josephson coupling and $\varphi_{\text{ref}}=\varphi_{\text{L}}-\varphi_{\text{R}}-2\pi\Phi/\Phi_{0}$ is the phase drop across the junction. The second term captures the indirect coupling of the SC leads through the SC island. We give the microscopic form of the Josephson coupling $J$ in  \cite{bib:supplemental}. Here, it suffices to remark that $J\neq0$ provided that $|\Gamma^{\text{L}}_{12}\Gamma^{\text{R}}_{12}|\neq0$ where $\Gamma^{m}_{ij}
\equiv
\pi\nu_{m}
(
\lambda^{\da}_{mi}\lambda^{\ua}_{mj}
-
\lambda^{\ua}_{mi}\lambda^{\da}_{mj}
)$ is the hybridization between the $m$-SC and the MBSs $\gamma_{i}$, $\gamma_{j}$. Lastly, the phase drop in the second term of the effective Hamiltonian is $\varphi=\varphi_{\text{L}}-\varphi_{\text{R}}+\varphi_0$ where $\varphi_0$ is an anomalous phase shift that arises from the complex tunnelling amplitudes $\lambda^{s}_{mi}$.

\begin{figure}[!t] \centering
\includegraphics[width=1\linewidth] {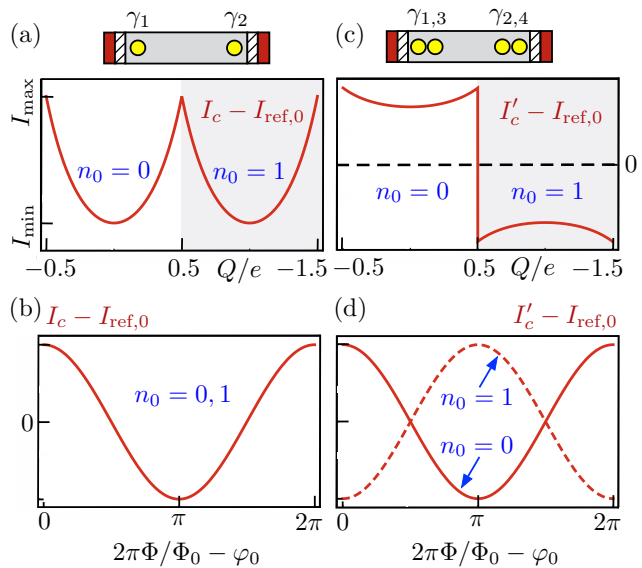}
\caption{(Color online)
(a) Schematic plot of the critical supercurrent $I_{c}$ passing through a MST versus the island-gate charge $Q$.The supercurrent is $1e$-periodic in the island-gate charge, not $2e$-periodic as for a trivial CPT. 
(b) Schematic plot of the critical supercurrent $I'_{c}$ versus the magnetic flux $\Phi$ through the SQUID loop of the MST. The critical supercurrent is independent of the total fermion parity $(-1)^{n_0}$ of the SC island. 
(c) Same as (a) but for an AST. The magnitude of the critical supercurrent is still $1e$-periodic in the island-gate charge. Yet, the sign of the critical supercurrent reverses when the fermion parity $(-1)^{n_0}$ of the SC island changes between even and odd. 
(d) Same as (b) but for an AST.  The critical supercurrent exhibits a sign reversal when the fermion parity $(-1)^{n_0}$ of the SC island switches between even and odd. 
}\label{fig:2}
\end{figure}

The effective Hamiltonian of Eq.~\eqref{Heff1} is our first main finding. The resulting supercurrent is given by 
\begin{equation}
I=I_{\text{ref},0}\sin\varphi_{\text{ref}}
+I_{0}\sin\varphi,
\label{Current}
\end{equation}
where $I_{\text{ref},0}=2eJ_{\text{ref}}/\hbar$ and $I_{0}=2eJ/\hbar$. The current-phase relation of Eq.~\eqref{Current} is measurable through the flux-dependence the critical supercurrent, $I_{c}=\text{max}_{\varphi}[I]$. For a highly-asymmetric SQUID, $I_{\text{ref},0}\gg I_{0}$, we have,
\begin{equation}
I_{c}=I_{\text{ref},0}+I_{0}\cos\left(\frac{2\pi\Phi}{\Phi_{0}}+\varphi_0\right).
\label{Eq11}
\end{equation}
Notably, this expression for the critical supercurrent depends on two tuning parameters: 

The first tuning parameter is the \textit{island-gate charge} $Q$ entering as a result of virtual transitions to excited charge states, $I_{0}=I_{0}(Q)$.  
In Fig.~\ref{fig:2}(a), we depict this dependence schematically for the case of zero energy splitting between the MBSs, $\varepsilon_{12}=0$. Notably, the critical supercurrent is $1e$-periodic in the gate charge $Q$ and independent of the total fermion parity $(-1)^{n_0}$ on the SC island. This $1e$-periodicity arises because the replacements $Q\rightarrow Q+e$, $n_{0}\rightarrow n_{0}+1$ leave the charging energy and, hence, the Josephson coupling $J$ invariant. In the next section, we will show that, interestingly, this $1e$-periodicity of the critical supercurrent will not carry over from the MST to the AST. 
At this point, two further remarks are in order:

 (i) The $1e$-periodicity of the critical supercurrent constitutes a sharp deviation from the $2e$-periodicity which, in the absence of quasiparticle poisoning, appears for a trivial CPT. Also, unlike in a trivial CPT, no above-gap quasiparticles need to be accessible in the SC island for a finite supercurrent. 
 
(ii) Any finite MBS energy splitting, $\varepsilon_{12}\neq0$, lifts the $1e$-periodicity of the supercurrent because the Josephson couplings for even and odd charge states acquire different energy denominators \cite{bib:supplemental}. A practical consequence is that we can use the deviation from perfect $1e$-periodicity as a tool for a qualitative estimate of the MBS energy splitting $\varepsilon_{12}$.

Returning to Eq.~\eqref{Eq11}, the second tuning parameter in the critical supercurrent is the \textit{magnetic flux} $\Phi$ threading the SQUID loop. 
In Fig.~\ref{fig:2}(b), we schematically plot the magnetic flux-dependence of the critical supercurrent. 
We again find that the current-phase relation is independent of the total fermion parity $(-1)^{n_0}$ on the SC island. 
This feature will not carry over from the MST to the AST. 

Finally, we note that the observation of a finite supercurrent across the SC island requires finite local and non-local couplings between the SC leads and the MBSs. As the non-local couplings are \textit{exponentially suppressed} in the length $d$ of the SC island, one might question if the above-described features of the supercurrent are measurable. Fortunately, measuring small currents does not pose an experimental challenge but requires only prolonged measurement times. Indeed, for a nanowire CPT subject to a magnetic field recent experiments observed a transition from a $2e$- to a $1e$-peak spacing in the magnitude of the switching current versus island-gate charge \cite{bib:Veen2018}. We suggest that such transition can arise as a result of nearly-zero-energy MBSs or, as we will see in the next section of our proposal, nearly-zero-energy ABSs.

{\it Supercurrent in the AST regime.}
We now turn our attention to the supercurrent for the AST where two ABSs, or equivalently four MBSs $\gamma_{1},\gamma_{2},\gamma_{3},\gamma_{4}$, mediate the Cooper pair transport between the SC leads. We focus on the limit of nearly-zero-energy ABSs, $\varepsilon_{ij}\ll\Delta,U$. This is the only limit that needs to be distinguished from the MST for qubit applications \cite{bib:Bravyi2010,bib:Vijay2015,bib:Vijay2016,bib:Landau2016,bib:Plugge2016,bib:Hoffman2016,bib:Vijay2016_2,bib:Plugge2017,bib:Karzig2016,bib:Schrade2018_2,bib:Schrade2018_3}.

As a first step, we note that the local couplings $\lambda^{s}_{\text{L}1(3)},\lambda^{s}_{\text{R}2(4)}$ between the ABSs and the lead electrons induce the dominant contribution to the supercurrent. As a result, we expect the supercurrent for the AST to be \textit{exponentially larger} compared to the MST where a finite supercurrent always requires finite non-local couplings. 
Notably, non-local couplings are also present for the AST but produce a contribution that is considerably weaker and, as we will argue, do not alter our findings qualitatively. For now, we set $\lambda^{s}_{\text{L}2(4)}=\lambda^{s}_{\text{R}1(3)}=0$.

Next, we notice that, for the same as reasons as in the MST regime, the second-order sequences in the tunnelling amplitudes $\lambda^{s}_{mi}$ do not contribute to the supercurrent. Consequently, the lowest-order contribution only appears in fourth-order of perturbation theory. In such a fourth-order sequence, a Cooper pair moves between the two SC leads by tunnelling in and out of the two ABSs, or equivalently the four MBSs, at the ends of the SC island. Notably, such sequences involve all four MBSs on the SC island, $\propto\gamma_{1}\gamma_{2}\gamma_{3}\gamma_{4}$, and, hence, depends on the total fermion parity of the SC island through Eq.~\eqref{Constraint}. 

We again compute the amplitudes of the relevant sequences perturbatively in the weak-coupling limit and summarize our findings by an effective Hamiltonian,
\begin{equation}
H'_{\text{eff}}=
-
J_{\text{ref}}\cos\varphi_{\text{ref}}
-
J'
(\gamma_{1}\gamma_{2}\gamma_{3}\gamma_{4})
\cos\varphi'\label{Heff2}.
\end{equation}
Here, for the Josephson junction which indirectly couples the SC leads via the SC island, we have introduced the phase drop
$\varphi'=\varphi_{\text{L}}-\varphi_{\text{R}}+\varphi'_0$ where $\varphi'_0$ is an anomalous phase shift that results because the tunnelling amplitudes $\lambda^{s}_{mi}$ are complex numbers. We give the microscopic form of the Josephson coupling $J'$ in \cite{bib:supplemental}. Here, we only note that $J'\neq0$ as long as $|\Gamma^{\text{L}}_{13}\Gamma^{\text{R}}_{24}|\neq0$. 

The effective Hamiltonian of Eq.~\eqref{Heff2} is our second main finding. The resulting supercurrent is 
\begin{equation}
\label{Current2}
I'=I_{\text{ref},0}\sin\varphi_{\text{ref}}+(-1)^{n_0}I'_{0}\sin\varphi',
\end{equation}
where $I'_{0}=2eJ'/\hbar$. We measure this current-phase relation of the supercurrent with the critical current through a highly asymmetric SQUID, $I_{\text{ref},0}\gg I'_{0}$. We have,
\begin{equation}
I'_{c}=I_{\text{ref},0}+(-1)^{n_0}I'_{0}\cos\left(\frac{2\pi\Phi}{\Phi_{0}}+\varphi_0\right).
\end{equation}
The critical supercurrent depends on two parameters: 

The first dependence is on the \textit{island-gate charge} $Q$ which we depict schematically in Fig.~\ref{fig:2}(c) for zero-energy ABSs, $\varepsilon_{ij}=0$. Here, the magnitude of the critical supercurrent is still $1e$-periodic in the gate charge $Q$. This behavior is identical to the MST regime and, therefore, does not allow us to make a distinction between nearly-zero-energy ABSs and MBSs. However, because the supercurrent depends on the total fermion parity $(-1)^{n_0}$ of the SC island, the sign of the critical supercurrent reverses when we tune the gate charge from $Q$ to $Q+e$. This sign reversal is also visible in the dependence on the the \textit{magnetic flux} $\Phi$, see Fig.~\ref{fig:2}(d). 
Most crucially, this sign reversal did not show up for a MST. Hence, it constitutes a distinctive feature by which we can distinguish the MST from the AST.

Before closing, we remark that in our calculations for the AST the supercurrent across the SC island only involves contributions that are parity-dependent. This is an outcome of our assumption of purely local couplings between the ABSs on the SC islands and the fermions in the SC leads. If we admit non-local couplings, as for the MST, parity-independent contributions will appear. These parity-independent contributions occur when a Cooper pair moves between the SC leads by tunnelling in and out via the same ABSs or, equivalently, the same two MBSs. However, it is important to note that these non-local contributions are significantly smaller in magnitude compared to the local contributions. As a result, it is not conceivable that the non-local contributions overwhelm the sign reversal of the supercurrent that arises due to the local contributions.

{\it Conclusions.}
We have shown that the magnitude of the supercurrent through the MST or the AST exhibits a $1e$-periodicity in the island-gate charge. This feature is unlike the trivial CPT where, in the absence of quasiparticle poisoning, the supercurrent is $2e$-periodic in the island-gate charge. Moreover, we have demonstrated that when tuning the island-gate charge between even and odd charge ground states the supercurrent reverses its sign for the AST. For the MST we find no such sign reversal. This peculiarity may help to clarify the nature of zero-energy bound states in TSC candidate systems and should be the first step towards Majorana qubit applications.

{\it Acknowledgments.}
We would like to thank Patrick A. Lee and Morten Kjaergaard for helpful discussions.
C.S. was supported by the Swiss SNF under Project 174980. L.F. and C.S. were supported by DOE Office of Basic Energy Sciences, Division of Materials Sciences and Engineering under Award $\text{DE-SC0010526}$.

\begin{widetext}

\newpage

\onecolumngrid

\bigskip

\begin{center}
\large{\bf Supplemental Material to `Andreev or Majorana, Cooper finds out' \\}
\end{center}
\begin{center}
Constantin Schrade and Liang Fu
\\
{\it Department of Physics, Massachusetts Institute of Technology, 77 Massachusetts Ave., Cambridge, MA 02139}
\end{center}

In the Supplemental Material, we provide the microscopic form of the Josephson couplings and anomalous phase shifts which appear in the effective Hamiltonians for the Majorana superconducting transistor and the Andreev superconducting transistor.  

\section{Effective Hamiltonian for the Majorana superconducting transistor}
In this first section of the Supplemental Material, we give the microscopic form of the MST-effective Hamiltonian when $0\leq\varepsilon_{12}\ll\Delta,U$. As an initial step, we recall from Eq.~(9) of the main text that the effective Hamiltonian is of the form,
\begin{equation}
H_{\text{eff}}=
-
J_{\text{ref}}\cos\varphi_{\text{ref}}
-
J
\cos\varphi.
\end{equation}
The expressions for the phase drop $\varphi_{\text{ref}}$ and the Josephson coupling $J_{\text{ref}}$ across the Josephson junction which directly couples the SC leads were already given in the main text. Here, we focus on the Josephson junction that indirectly connects the SC leads via the SC island. 
First, we find that the phase drop is given by,
\begin{equation}
\varphi=\varphi_{\text{L}}-\varphi_{\text{R}}+\varphi_0
\quad \text{with} \quad \varphi_{0}=\arg[
(\Gamma^{\text{L}}_{12})^{*}\Gamma^{\text{R}}_{12}
].
\end{equation}
Second, the Josephson couplings reads $J=\sum_{s,s'=\pm}J_{ss'}$ with
\begin{align}
J_{ss'}
&=
\frac{16|\Gamma^{\text{L}}_{12}\Gamma^{\text{R}}_{12}|}{\pi^{2}\Delta}
\int^{\infty}_{1}
\frac
{
\mathrm{d}x \text{ }�� \mathrm{d}y\ (\delta_{ss'}-1)
}{
f(x)f(y)
\left[f(x)+f(y)\right]g_{s}(x)g_{s'}(y)}
-
\frac{16|\Gamma^{\text{L}}_{12}\Gamma^{\text{R}}_{12}|}{\pi^{2} h_{s}}
\left[
\int^{\infty}_{1}
\frac
{
\mathrm{d}x \text{ }��
}{
f(x)g_{s}(x)} 
\right]^{2} \delta_{ss'}
.
\end{align}
Here, for the case $\varepsilon_{12}=0$, we have introduced the functions, 
\begin{equation}
\begin{split}
f(x)&\equiv\sqrt{1+x^{2}},\quad g_{s}(x)\equiv f(x)+\frac{U_{C}(n_{0}+s)-U_{C}(n_{0})}{\Delta}
,\quad h_{s}\equiv U_{C}(n_{0}+2s)-U_{C}(n_0). 
\end{split}
\end{equation}
For the case $0<\varepsilon_{12}\ll\Delta,U$, we have the same expressions but with the replacement 
\begin{equation}
U_{C}(n) \rightarrow U_{C}(n)+(-1)^{n}\varepsilon_{12}.
\end{equation}
Notably, as long as $\varepsilon_{12}=0$, the supercurrent is $1e$-periodic in the island gate charge. This periodicity is lifted once $0<\varepsilon_{12}\ll\Delta,U$, see Fig.~3.

\section{Effective Hamiltonian for the Andreev superconducting transistor}
In this second section of the Supplemental Material, we present microscopic form of the effective Hamiltonian of a AST. For simplicity, we will only consider the case when $\varepsilon_{ij}=0$. First, we recall from Eq.~(12) of the main text that,
\begin{equation}
H'_{\text{eff}}=
-
J_{\text{ref}}\cos\varphi_{\text{ref}}
-
(\gamma_{1}\gamma_{2}\gamma_{3}\gamma_{4})
J'
\cos\varphi'.
\end{equation}
Here, the phase drop across the Josephson junction with the SC island is given by
\begin{equation}
\varphi=\varphi_{\text{L}}-\varphi_{\text{R}}+\varphi'_0
\quad \text{with} \quad\varphi'_{0}=\arg[
(\Gamma^{\text{L}}_{13})^{*}\Gamma^{\text{R}}_{24}
].
\end{equation}
Second, we find for the Josephson couplings that $J'=\sum_{s,s'=\pm}J'_{ss'}$ with
\begin{align}
J'_{ss'}
&=
-\frac{16|\Gamma^{\text{L}}_{13}\Gamma^{\text{R}}_{24}|}{\pi^{2}\Delta}
\int^{\infty}_{1}
\frac
{
\mathrm{d}x \text{ }�� \mathrm{d}y
}{
f(x)f(y)
\left[f(x)+f(y)\right]g_{s}(x)g_{s'}(y)}-
\frac{16|\Gamma^{\text{L}}_{13}\Gamma^{\text{R}}_{24}|}{\pi^{2} h_{s}}
\left[
\int^{\infty}_{1}
\frac
{
\mathrm{d}x \text{ }��
}{
f(x)g_{s}(x)} 
\right]^{2} \delta_{ss'}
.
\end{align}

\begin{figure}[!b] \centering
\includegraphics[width=0.7\linewidth] {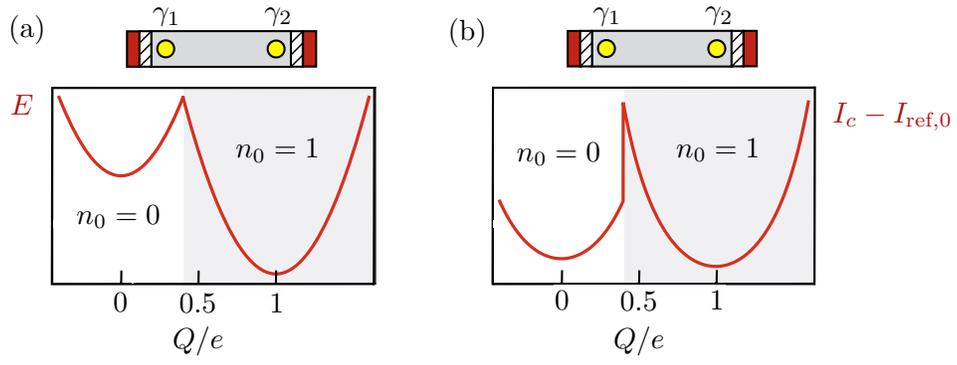}
\caption{(Color online)
(a) Schematic plot of the energy $E$ of the island-charge ground state versus gate charge $Q$ for $0<\varepsilon_{12}\ll\Delta,U$. The $1e$-periodicity in $Q$, which is present as long as  $\varepsilon_{12}=0$, lifts once $\varepsilon_{12}>0$. 
(b) Schematic plot of the critical current $I_{c}$ passing through the MST versus gate charge $Q$ for $0<\varepsilon_{12}\ll\Delta,U$. Similar to (a), the $1e$-periodicity in $Q$, which is present for $\varepsilon_{12}=0$, lifts once 
$\varepsilon_{12}>0$. 
}\label{fig:1_SM}
\end{figure}

\end{widetext}

\end{document}